# A Nearly Optimal Construction of Flash Codes

**Hessam Mahdavifar, Paul H. Siegel, Alexander Vardy, Jack K. Wolf,** and **Eitan Yaakobi**
Department of Electrical Engineering, University of California San Diego, La Jolla, CA 92093, USA
{hessam,psiegel,avardy,jwolf,eyaakobi}@ucsd.edu

*Abstract*— Flash memory is a non-volatile computer memory comprised of blocks of cells, wherein each cell can take on $q$ different values or *levels*. While increasing the cell level is easy, reducing the level of a cell can be accomplished only by erasing an entire block. Since block erasures are highly undesirable, coding schemes — known as *floating codes* or *flash codes* — have been designed in order to maximize the number of times that information stored in a flash memory can be written (and re-written) prior to incurring a block erasure. An $(n,k,t)_q$ flash code $\mathbb{C}$ is a coding scheme for storing $k$ information bits in $n$ cells in such a way that any sequence of up to $t$ writes (where a *write* is a transition $0 \to 1$ or $1 \to 0$ in any one of the $k$ bits) can be accommodated without a block erasure. The total number of available level transitions in $n$ cells is $n(q-1)$, and the *write deficiency* of $\mathbb{C}$, defined as $\delta(\mathbb{C}) = n(q-1) - t$, is a measure of how close the code comes to perfectly utilizing all these transitions. For $k > 6$ and large $n$, the best previously known construction of flash codes achieves a write deficiency of $O(qk^2)$. On the other hand, the best known lower bound on write deficiency is $\Omega(qk)$. In this paper, we present a new construction of flash codes that approaches this lower bound to within a factor *logarithmic* in $k$. To this end, we first improve upon the so-called "indexed" flash codes, due to Jiang and Bruck, by eliminating the need for index cells in the Jiang-Bruck construction. Next, we further increase the number of writes by introducing a new multi-stage (recursive) indexing scheme. We then show that the write deficiency of the resulting flash codes is $O(qk \log k)$ if $q \geqslant \log_2 k$, and at most $O(k \log^2 k)$ otherwise.

## I. Introduction

Flash memories are, by far, the most important type of nonvolatile computer memory in use today. Flash devices are employed widely in mobile, embedded, and mass-storage applications, and the growth in this sector continues at a staggering pace.

A flash memory consists of an array of floating-gate **cells**, organized into **blocks** (a typical block contains $2^{18}$ to $2^{20}$ cells). The level or "state" of a cell is a function of the amount of charge (electrons) trapped within it. In *multilevel flash cells*, voltage is quantized to $q$ discrete threshold values; consequently the level of each cell can be modeled as an integer in the range $0, 1, \ldots, q-1$. The parameter $q$ itself ranges from $q = 2$ (the conventional two-state case) up to $q = 256$. The most conspicuous property of flash-storage technology is its inherent asymmetry between cell programming (charge placement) and cell erasing (charge removal). While adding charge to a single cell is a fast and simple operation, removing charge from a cell is very difficult. In fact, flash technology does not allow a single cell to be erased — rather, only entire blocks can be erased. Such **block erasures** are not only time-consuming, but also degrade the physical quality of the memory. For example, a typical block in a multilevel flash memory can tolerate only about $10^4$ erasures before it becomes unusable. Therefore, it is of importance to design coding schemes that maximize the number of times information stored in a flash memory can be written (and re-written) prior to incurring a block erasure.

Such coding schemes — known as **floating codes** or **flash codes** — were first introduced in [3] two years ago. Since then, a few more papers on this subject have appeared in the literature [2,4,6,8]. It should be pointed out, however, that flash codes may be regarded as a generalization of codes for write-once memories [1,7], that were studied since the early 1980s.

An $(n,k,t)_q$ flash code $\mathbb{C}$ is a coding scheme for storing $k$ information bits in $n$ flash-memory cells, with $q$ levels each, in such a way that any sequence of up to $t$ writes can be accommodated without incurring a block erasure. In the literature on flash codes, a *write* is always a bit-write — that is, a change $0 \to 1$ or $1 \to 0$ in the value of one of the $k$ information bits. Observe that in order to accommodate such a write, at least one of the $n$ cells must transition from a lower level to a higher level (since a cell's level, determined by its charge, can only increase). On the other hand, the total number of available level transitions in $n$ flash cells is $n(q-1)$. Thus, throughout this paper, we characterize the performance of a flash code $\mathbb{C}$ in terms of its **write deficiency**, defined as $\delta(\mathbb{C}) = n(q-1) - t$. According to the foregoing discussion, $\delta(\mathbb{C})$ is a measure of how close $\mathbb{C}$ comes to perfectly utilizing all the available cell-level transitions: exactly one per write. The primary goal in designing flash codes can thus be expressed as *minimizing deficiency*.

What is the smallest possible write deficiency $\delta_q(n,k)$ for an $(n,k,t)_q$ flash code, and how does it behave asymptotically as the code parameters $k$ and $n$ get large? The best-known lower bound, due to Jiang, Bohossian, and Bruck [3], asserts that

$$\delta_q(n,k) \;\geqslant\; \tfrac{1}{2}(q-1) \min\{n, k-1\} \qquad (1)$$

How closely can this bound be approached by code constructions? It appears that the answer to this question depends on the relationship between $k$ and $n$. In this paper, we are concerned mainly with the case where both $k$ and $n$ are large, and $n$ is much larger than $k$ (in particular, $n \geqslant k^2$). In Section V, we briefly consider the case $k/n = $ const. At the other end of the spectrum, the case $k > n$ has been recently studied in [5].

The first construction of flash codes for large $k$ was reported by Jiang and Bruck [4]. In this construction, the $k$ information bits are partitioned into $m_1 = k/k'$ subsets of $k'$ bits each (with $k' \leqslant 6$) while the memory cells are subdivided into $m_2 \geqslant m_1$ groups of $n'$ cells each. Additional memory cells (called **index cells**) are set aside to indicate for each subset of $k'$ bits which group of $n'$ memory cells is used to store them. The deficiency of the resulting codes is at least $O(\sqrt{qn})$. Note that for $n \geqslant k$, the lower bound on write deficiency in (1) behaves as $\Omega(qk)$, and thus does not depend on $n$. Consequently, the gap between the Jiang-Bruck construction [4] and the lower bound could be arbitrarily large, especially when $n$ is much larger than $k$.

Recently, we have proposed in [8] a completely different construction of flash codes. These codes are based upon representing the $n$ memory cells as a high-dimensional array, and achieve a write deficiency of $O(qk^2)$. Crucially, the deficiency of these codes does *not* depend on $n$. Nevertheless, there is still a significant gap between $O(qk^2)$ — which is the best currently known result — and the lower bound of $\Omega(qk)$.

In this paper, we present a new construction of flash codes which reduces the gap between the upper and lower bounds on write deficiency to a factor that is *logarithmic* in the number of information bits $k$. This result is arrived at in several stages. As a starting point, we use the "indexed" flash codes of Jiang and Bruck [4]. In Section III, we develop new encoding and decoding procedures for such codes that eliminate the need for index cells in the Jiang-Bruck construction [4]. The write deficiency achieved thereby is $O(qk^2)$, which coincides with the main result of [8]. When the encoding procedure developed in Section III reaches its limit, there are still potentially numerous unused cell-level transitions. In Section IV, we show how to take advantage of these transitions in order to accommodate even more writes. To this end, we introduce a new indexing scheme, which is invoked only after the encoding method of Section III is exhausted. Thereupon, we extend this idea recursively, through $\lceil \log_2 k \rceil$ different indexing stages. This leads to the main result of this paper, established in Theorem 3, namely

$$\Omega(qk) \leqslant \delta_q(n,k) \leqslant O(\max\{q, \log_2 k\} \, k \log k) \quad (2)$$

for all $n \geqslant k^2$, where the upper bound is achieved constructively by the flash codes described in Section IV.

Finally, in Section V, we present and briefly discuss constructions of flash codes for the case where the number of memory cells $n$ is not significantly larger than the number of bits $k$.

## II. PRELIMINARIES

Let us now give a precise definition of flash codes that were introduced less formally in the previous section. We use $\{0,1\}^k$ to denote the set of binary vectors of length $k$, and refer to the elements of this set as ***information vectors***. The set of possible levels for each cell is denoted by $\mathcal{A}_q = \{0, 1, \ldots, q-1\}$ and thought of as a subset of the integers. The $q^n$ vectors of length $n$ over $\mathcal{A}_q$ are called ***cell-state vectors***. With this notation, any flash code $\mathbb{C}$ can be specified in terms of two functions: an encoding map $\mathcal{E}$ and a decoding map $\mathcal{D}$. The ***decoding map*** $\mathcal{D} \colon \mathcal{A}_q^n \to \{0,1\}^k$ indicates for each cell-state vector $\boldsymbol{x} \in \mathcal{A}_q^n$ the corresponding information vector. In turn, the ***encoding map*** $\mathcal{E} \colon \{0,1,\ldots,k-1\} \times \mathcal{A}_q^n \to \mathcal{A}_q^n \cup \{\mathsf{E}\}$ assigns to every index $i$ and cell-state vector $\boldsymbol{x} \in \mathcal{A}_q^n$, another cell-state vector $\boldsymbol{y} = \mathcal{E}(i, \boldsymbol{x})$ such that $y_j \geqslant x_j$ for all $j$ and $\mathcal{D}(\boldsymbol{y})$ differs from $\mathcal{D}(\boldsymbol{x})$ only in the $i$-th position. If no such $\boldsymbol{y} \in \mathcal{A}_q^n$ exists, then $\mathcal{E}(i, \boldsymbol{x}) = \mathsf{E}$ indicating that block erasure is required. To bootstrap the encoding process, we assume that the initial state of the $n$ memory cells is $(0,0,\ldots,0)$. Henceforth, iteratively applying the encoding map, we can determine how *any sequence* of transitions $0 \to 1$ or $1 \to 0$ in the $k$ information bits maps into a sequence of cell-state vectors, eventually terminated by the block erasure. This leads to the following definition.

**Definition.** *An $(n,k)_q$ flash code $\mathbb{C}(\mathcal{D}, \mathcal{E})$ **guarantees** $t$ **writes** if for all sequences of up to $t$ transitions $0 \to 1$ or $1 \to 0$ in the $k$ information bits, the encoding map $\mathcal{E}$ does not produce the block erasure symbol $\mathsf{E}$. If so, we say that $\mathbb{C}$ is an $(n,k,t)_q$ code, and define the **deficiency of** $\mathbb{C}$ as $\delta(\mathbb{C}) = n(q-1) - t$.*

In addition to this definition, we will also use the following terminology. Given a vector $\boldsymbol{x} = (x_1, x_2, \ldots, x_m)$ over $\mathcal{A}_q$, we define its ***weight*** as $\mathrm{wt}(\boldsymbol{x}) = x_1 + x_1 + \cdots + x_m$ (where the addition is over the integers), and its ***parity*** as $\mathrm{wt}(\boldsymbol{x}) \bmod 2$.

## III. INDEX-LESS INDEXED FLASH CODES

Our point of departure are the so-called *indexed flash codes*, due to Jiang and Bruck [4], that were briefly described in Section I. In this section, we eliminate the need for index cells — and, thus, the overhead associated with these cells — in the Jiang-Bruck construction [4]. This is achieved by "encoding" the indices into the order in which the cell levels are increased.

As in [4], we partition the $n$ memory cells into $m$ groups of $n'$ cells each. However, while in [4] the value of $n'$ is more or less arbitrary, in our construction $n' = k$. We henceforth refer to such groups of $n' = k$ cells as ***blocks*** (though they are not related to the *physical blocks* of floating-gate cells which comprise the flash memory). We will furthermore use, throughout this paper, the following terminology. We say that:
▶ a block is ***full*** if all its cells are at level $q-1$;
▶ a block is ***empty*** if all its cells are at level zero;
▶ a block is ***active*** if it is neither full nor empty;
▶ a block is ***live*** if it is not full (either active or empty).

In our construction, each block represents *exactly one bit*. This implies that the total number of blocks, given by $m = \lfloor n/k \rfloor$, must be at least $k$, which in turn implies $n \geqslant k^2$. If $n$ is not divisible by $k$, the remaining cells are simply left unused. Finally, we also assume that either $k$ is even or $q$ is odd. If this is not the case, we can invoke the same construction with $k$ replaced by $k+1$ (and the last bit permanently set to zero).

The key idea is that each block is used to encode not only the current value of the bit that it represents, but also *which* of the $k$ bits it represents. The value of the bit is simply the parity of the block. The index of the bit is encoded in the *order* in which the levels of the $k$ cells are increased. For example, if the block stores the $i$-th bit, first the level of the $i$-th cell in the block is increased from $0$ to $q-1$ in response to the transitions $0 \to 1$ and $1 \to 0$ in the bit value. Then, the same procedure is applied to the $(i+1)$-st cell, the $(i+2)$-nd cell, and so on, with the indices $i+1, i+2, \ldots$ interpreted cyclically (modulo $k$). This process is illustrated in the following example.

**Example.** Suppose that $k = 4$ and $q = 3$. If a block represents the first bit, then its cell levels will transition from $(0,0,0,0)$ to $(2,2,2,2)$ in the following order:

$$(0000) \to (1000) \to (2000) \to (2100) \to (2200)$$
$$\to (2210) \to (2220) \to (2221) \to (2222)$$

On the other hand, for a block that represents the second bit, the corresponding cell-writing order is given by:

$$(0000) \to (0100) \to (0200) \to (0210) \to (0220)$$
$$\to (0221) \to (0222) \to (1222) \to (2222)$$

The cell-writing orders for blocks that represent the third and fourth bits are given, respectively, by

$$(0000) \to (0010) \to (0020) \to (0021) \to (0022)$$
$$\to (1022) \to (2022) \to (2122) \to (2222)$$

and
$$(0000) \to (0001) \to (0002) \to (1002) \to (2002)$$
$$\to (2102) \to (2202) \to (2212) \to (2222)$$

Note that, unless a block is full, it is always possible to determine which cell was written first and, consequently, which of the $k = 4$ bits this block represents. □

We now provide a precise specification of an $(n,k)_q$ flash code $\mathbb{C}$ based upon this idea, in terms of a decoding map $\mathcal{D}_0$ and an encoding map $\mathcal{E}_0$. In what follows, these maps are described algorithmically, using (C-like) pseudo-code notation.

**Decoding map $\mathcal{D}_0$:** The input to this map is a cell-state vector $x = (x_1|x_2|\cdots|x_m)$, partitioned into $m$ blocks of $k$ cells. The output is the information vector $(v_0, v_1, \ldots, v_{k-1})$.

```
(v_0, v_1, ..., v_{k-1}) = (0, 0, ..., 0);
for (j = 1; j ≤ m; j = j+1)
if (active(x_j))
{ i = read_index(x_j); v_i = parity(x_j); }
```

**Encoding map $\mathcal{E}_0$:** The input to this map is a cell-state vector $x = (x_1|x_2|\cdots|x_m)$, partitioned into $m$ blocks of $k$ cells, and an index $i$ of the bit that has changed. Its output is either a cell-state vector $y = (y_1|y_2|\cdots|y_m)$ or the erasure symbol E.

```
(y_1|y_2|···|y_m) = (x_1|x_2|···|x_m);
for (j = 1; j ≤ m; j = j+1)
if (active(x_j) ∧ (read_index(x_j) == i))
{ write(y_j); break; }

if (j == m+1) // active block not found
for (j = 1; j ≤ m; j = j+1)
if (empty(x_j)) { write_new(i, y_j); break;}

if (j == m+1) // no empty blocks remain
return E;
```

To complete the specification of the flash code $\mathbb{C}(\mathcal{D}_0, \mathcal{E}_0)$, let us elaborate upon all the functions used in the pseudo-code above. The functions `active(x)`, respectively `empty(x)`, simply determine whether the given block is active, respectively empty. The function `parity(x)` computes the parity of $x$, defined in Section II. Note that the parity of a full block is always zero (since $k(q-1)$ is even, by assumption). The function `read_index(x)` computes the bit-index encoded in an active block $x = (x_0, x_1, \ldots, x_{k-1})$. This can be done as follows. Find all the zero cells in $x$. Note that these cells always form one cyclically contiguous run, say $x_j, x_{j+1}, \ldots, x_{j+r}$ (where the indices are modulo $k$). Then the index of the corresponding bit is $i = j + r + 1 \pmod{k}$. If there are no zeros in $x$, there must be exactly one cell, say $x_j$, whose level is strictly less than $q-1$. In this case the bit-index is $i = j + 1 \pmod{k}$. The function `write(y)` proceeds along similar lines. Find the single cyclically contiguous run of zeros in $(y_0, y_1, \ldots, y_{k-1})$, say $y_j, y_{j+1}, \ldots, y_{j+r}$. If $y_{j-1} < q-1$, increase $y_{j-1}$ by one; otherwise set $y_j = 1$. If there are no zeros in $y$, find the unique cell $y_j$ such that $y_j < q-1$ and increase its level by one. Finally, the function `write_new(i, y)` simply sets $y_i = 1$.

**Theorem 1.** *The write deficiency of the flash code $\mathbb{C}(\mathcal{D}_0, \mathcal{E}_0)$ described above is at most*

$$(k-1)\big((k+1)(q-1) - 1\big) = O(qk^2) \quad (3)$$

*Proof.* Note that at each instance, at most $k$ of the $m$ blocks are active. The encoding map $\mathcal{E}_0(i, x)$ produces the symbol E when there are no more empty blocks, and none of the active blocks represents the $i$-th bit. In the worst case, this may occur when there are $k-1$ active blocks, each using just one cell level. This contributes $(k-1)(k(q-1) - 1)$ unused cell levels. In addition, there are at most $k-1$ cells that are unused due to the partition into $m = \lfloor n/k \rfloor$ blocks of exactly $k$ cells. These contribute at most $(k-1)(q-1)$ unused cell levels. ∎

## IV. NEARLY OPTIMAL CONSTRUCTION

It is apparent from the proof of Theorem 1 that the deficiency of the flash code $\mathbb{C}(\mathcal{D}_0, \mathcal{E}_0)$, constructed in Section III, is due primarily to the following: when writing stops, there are still potentially numerous unused cell levels. The key idea developed in this section is to *continue writing* after the encoding map $\mathcal{E}_0$ produces the erasure symbol E, utilizing those cell levels that are left unused by $\mathcal{E}_0$. Obviously, it is *not* possible to continue writing using the same encoding and decoding maps. However, it may be possible to do so if, at the point when $\mathcal{E}_0$ produces the erasure symbol E, we switch to a *different encoding procedure*, say $\mathcal{E}_1$. In fact, this idea can be applied iteratively: once $\mathcal{E}_1$ reaches its limit, we will transition to another encoding map $\mathcal{E}_2$, then yet another map $\mathcal{E}_3$, and so on.

Assuming that $k \equiv 0 \pmod{4}$, here is one way to continue writing after the encoding map $\mathcal{E}_0$ has been exhausted. When $\mathcal{E}_0$ produces the erasure symbol E, we say that the *first stage* of encoding is over and transition to the *second stage*, as follows. First, we re-examine the cell-state vector $x = (x_1|x_2|\cdots|x_m)$ and re-partition it into $2m = 2\lfloor n/k \rfloor$ blocks of $k/2$ cells each. Most of these smaller blocks will be already full, but we may find some $m_1$ of them that are either empty or active (live). Observe that $m_1 \leq 2(k-1)$ since at the end of the first stage, there are at most $k-1$ active blocks of $k$ cells, and each of them produces at most two live (non-full) blocks of $k/2$ cells.

If $m_1 \geq k$, we can continue writing as follows. Once again, each of the $m_1$ blocks will represent exactly one bit; as before, the value of this bit is determined by the parity of the block. As part of the transition from the first stage to the second stage, we record the current information vector $(v_0, v_1, \ldots, v_{k-1})$ in the first $k$ of the $m_1$ live blocks, say $x_1, x_2, \ldots, x_k$. To this end, whenever $\texttt{parity}(x_i) \neq v_{i-1}$, we increase the level of one of the cells in $x_i$ by one; otherwise, we leave $x_i$ as is.

Since the blocks now have $k/2$ cells rather than $k$ cells, it is no longer possible to encode in each block *which* of the $k$ information bits it represents. Therefore, we set aside for this purpose $2(k-1)\lceil \log_q(k+2) \rceil$ index cells (that are not used during the first stage). These cells are partitioned into $2(k-1)$ blocks of $\mu = \lceil \log_q(k+2) \rceil$ cells each, which we call **index blocks**. Henceforth, it will be convenient to refer to the blocks of $k/2$ cells as **parity blocks**, in order to distinguish them from the index blocks. Initially, the first $k$ index blocks $u_1, u_2, \ldots, u_k$ are set so that $u_i = i$ (in the base-$q$ number system), which reflects the fact that the information bits $v_0, v_1, \ldots, v_{k-1}$ are stored (in that order) in the first $k$ live parity blocks. The next $m_1 - k$ index blocks are set to $(0, 0, \ldots, 0)$, thereby indicating that the corresponding (live) parity blocks are available to store information bits. The last $2(k-1) - m_1$ index blocks are set to $(q-1, q-1, \ldots, q-1)$ to indicate that the corresponding parity blocks are full (in fact, nonexistent). Finally, it is possible that in the process of enforcing $\texttt{parity}(x_i) = v_{i-1}$ for the first $k$ live parity blocks, some of these blocks become full (this happens iff $\text{wt}(x_i) = (k/2)(q-1) - 1$ and $v_i = 0$ at the end of the first stage, since $k/2$ is even by assumption). To account for this fact, we set the corresponding index blocks to $(q-1, q-1, \ldots, q-1)$. This completes the transition from the first stage to the second stage, which is invoked when the encoding map $\mathcal{E}_0$ produces the erasure symbol E.

Let us now summarize the foregoing discussion by giving a concise algorithmic description of the transition procedure.

**Transition procedure $\mathcal{T}_1$:** Partition the memory into $2\lfloor n/k \rfloor$ parity blocks of $k/2$ cells, and identify the $m_1 \leqslant 2(k-1)$ parity blocks $x_1, x_2, \ldots, x_{m_1}$ that are not full. If $m_1 < k$, output the erasure symbol E and terminate. Otherwise, set the $2(k-1)$ index blocks $u_1, u_2, \ldots, u_{2k-2}$ as follows:

$$u_i = \begin{cases} i & \text{for } i = 1, 2, \ldots, k \\ 0 & \text{for } i = k+1, k+2, \ldots, m_1 \\ q^\mu - 1 & \text{for } i = m_1+1, m_1+2, \ldots, 2k-2 \end{cases} \quad (4)$$

where $\mu = \lceil \log_q(k+2) \rceil$ is the number of cells in each index block, then record the information vector $(v_0, v_1, \ldots, v_{k-1})$ in the first $k$ live parity blocks $x_1, x_2, \ldots, x_k$, as follows:

```
for (i = 1; i ⩽ k; i = i+1)
if (parity(x_i) ≠ v_{i-1})
{ increment(x_i); if (full(x_i)) u_i = q^μ − 1; }
```

The function **full(x)** determines whether the given block $x$ (which could be a parity block or an index block) is full. The function **increment(x)** increases by one the level of a cell (does not matter which) in the given live block.

During second-stage encoding and decoding, we will need to figure out for each active parity block $x$ which of the $k$ information bits it represents. To this end, we will have to find and read the index block $u$ that *corresponds* to $x$. How exactly is the correspondence between parity blocks and index blocks established? Note that, upon the completion of the transition procedure $\mathcal{T}_1$, there is the same number of live parity blocks and live index blocks; moreover, the $j$-th *live* index block corresponds to the $j$-th *live* parity block, for all $j$. The encoding procedure will make sure that this correspondence is preserved throughout the second stage: whenever a parity block becomes full, it will make the corresponding index block full as well.

We are now ready to present the encoding and decoding maps which are, again, specified in C-like pseudo-code notation.

**Decoding map $\mathcal{D}_1$:** The input to this map is a cell-state vector $x = (x_1|x_2|\cdots|x_{2m} \| u_1|u_2|\cdots|u_{2k-2})$, partitioned into $2m$ parity blocks, of $k/2$ cells each, and $2(k-1)$ index blocks. The output is the information vector $(v_0, v_1, \ldots, v_{k-1})$.

```
(v_0, v_1, ..., v_{k-1}) = (0, 0, ..., 0);
for (ℓ = j = 1; j ⩽ 2m; j = j+1)
{
    if (full(x_j)) continue;  // skip full blocks
    while (full(u_ℓ)) ℓ = ℓ+1; // skip full blocks
    i = u_ℓ;  ℓ = ℓ+1;
    if (i ≠ 0) v_{i-1} = parity(x_j);
}
```

Given an index $i$ of the bit that has changed, the encoding map $\mathcal{E}_1$ first tries to find an active parity block $x$ that represents the $i$-th information bit. If such a block is found, it is incremented and checked for getting full (in which case the corresponding index block is set to $q^\mu - 1$). If not, another live parity block is allocated to represent the $i$-th information bit. If no more live parity blocks are available, the erasure symbol E is returned.

**Encoding map $\mathcal{E}_1$:** The input to this map is a cell-state vector $x = (x_1|x_2|\cdots|x_{2m} \| u_1|u_2|\cdots|u_{2k-2})$, partitioned into $2m$ parity blocks and $2(k-1)$ index blocks, and an index $i$ of the information bit that changed. Its output is either a cell-state vector $y = (y_1|y_2|\cdots|y_{2m} \| u'_1|u'_2|\cdots|u'_{2k-2})$ or the symbol E.

```
(y_1|y_2|···|y_{2m}) = (x_1|x_2|···|x_{2m});
(u'_1|u'_2|···|u'_{2k-2}) = (u_1|u_2|···|u_{2k-2});
for (ℓ = j = 1; j ⩽ 2m; j = j+1)
{
    if (full(x_j)) continue;
    while (full(u_ℓ)) ℓ = ℓ+1;
    if (u_ℓ == i+1)
    {
        increment(y_j);
        if (full(y_j)) u'_ℓ = q^μ − 1;
        break;
    }
    else ℓ = ℓ+1;
}
if (j == 2m+1) // active block not found
for (ℓ = j = 1; j ⩽ 2m; j = j+1)
{
    if (full(x_j)) continue;
    while (full(u_ℓ)) ℓ = ℓ+1;
    if (u_ℓ == 0)
    {
        u'_ℓ = i+1;
        if (parity(x_j) ≠ v_i) increment(y_j);
        if (full(y_j)) u'_ℓ = q^μ − 1;
        break;
    }
    else ℓ = ℓ+1;
}
if (j == 2m+1) // no more available live blocks
return E;
```

Note that when the second encoding stage terminates, there are at most $k-1$ parity blocks that are not full, comprising at most $k(k-1)/2$ cells (at most $k(k-1)(q-1)/2$ cell-levels).

Once the maps $\mathcal{D}_1$ and $\mathcal{E}_1$ are understood, it becomes clear that the same approach can be applied iteratively. The resulting flash code $\mathbb{C}^*$ will proceed, sequentially, through $s$ different encoding stages $\mathcal{E}_0, \mathcal{E}_1, \ldots, \mathcal{E}_{s-1}$, where $s = \lceil \log_2 k \rceil$. In describing this code, we shall assume for the sake of simplicity that $k$ is a power of two, that is $k = 2^s$. If not, the same code can be used to store $2^s > k$ information bits, of which the last $2^s - k$ are set to zero. Note that this will not change the order of the resulting write deficiency.

To accommodate the encoding maps $\mathcal{E}_1, \mathcal{E}_2, \ldots, \mathcal{E}_{s-1}$, we set aside for *each map* a batch of $2(k-1)$ index blocks, with each index block consisting of $\mu = \lceil \log_q(k+2) \rceil$ cells. The transition procedure $\mathcal{T}_r$ which bridges between the encoding maps $\mathcal{E}_{r-1}$ and $\mathcal{E}_r$ (for some $r \in \{2, 3, \ldots, s-1\}$) is identical to the transition procedure $\mathcal{T}_1$, except for the following differences:

**D1.** The $r$-th batch of index blocks is used; and

**D2.** The parity blocks consist of $k/2^r$ cells each.

In addition to **D1** and **D2**, the decoding/encoding maps $\mathcal{D}_r$ and $\mathcal{E}_r$ differ from $\mathcal{D}_1$ and $\mathcal{E}_1$ in that "$2m$" should be replaced by "$2^r m$" throughout, where $m$ stands for $\lfloor n/k \rfloor$ as before. There are no other differences.

**Theorem 2.** *For $s = \lceil \log_2 k \rceil$, the write deficiency of the flash code $\mathbb{C}^*$ defined by the sequence of decoding/encoding maps $\mathcal{D}_0, \mathcal{D}_1, \ldots, \mathcal{D}_{s-1}$ and $\mathcal{E}_0, \mathcal{E}_1, \ldots, \mathcal{E}_{s-1}$ is $O(qk \log^2 k / \log q)$.*

*Proof.* We consider the worst-case scenario for the number of cell levels that are either unused or "wasted" in the overall encoding procedure. As before, there are at most $k-1$ cells that are unused due to the partition into $\lfloor n/k \rfloor$ blocks, of ex-actly $k$ cells each, at the very first encoding stage. These cells contribute at most $(q-1)(k-1)$ unused cell levels. The index blocks for the $s-1$ encoding maps $\mathcal{E}_1, \mathcal{E}_2, \ldots, \mathcal{E}_{s-1}$ contain $2(k-1)(s-1)\mu$ cells altogether, thereby wasting at most

$$2(q-1)(k-1)(s-1)\lceil \log_q(k+2)\rceil \;=\; O\!\left(\frac{qk\log^2 k}{\log q}\right) \quad (5)$$

cell levels. In each of the $s-1$ transition procedures, the situation `parity(x_i)` $\neq v_{i-1}$ can occur at most $k$ times, and each time it occurs a single cell level is wasted. Finally, as in Theorem 1, when the encoding process $\mathcal{E}_0, \mathcal{E}_1, \ldots, \mathcal{E}_{s-1}$ terminates there are at most $k-1$ parity blocks that are not full and, in the worst case, each of them uses just one cell level. However, now these parity blocks contain only $\lceil k/2^{s-1}\rceil = 2$ cells each, and thus contribute at most $(k-1)(2q-3)$ unused cell levels. Putting all of this together, we find that at most

$$(q-1)(k-1)\Big(2(s-1)\lceil \log_q(k+2)\rceil + 3\Big) + k(s-1) \quad (6)$$

cell levels are wasted or left unused. Clearly, this expression is dominated by (5), and thus bounded by $O(qk\log^2 k/\log q)$. ∎

For large $q$, the upper bound of $O(qk\log^2 k/\log q)$ on the deficiency of our scheme can be improved by using a more efficient "packaging" of index blocks in the flash memory. As before, we allocate a batch of $2(k-1)$ index blocks to each encoding stage except $\mathcal{E}_0$. But now, every index block will occupy $\mu' = \lceil \log_2(k+2)\rceil$ cells rather than $\mu = \lceil \log_q(k+2)\rceil$ cells, and the indices will be written in binary rather than in the base-$q$ number system. This allows index blocks that correspond to successive encoding stages to be "stacked on top of each other" in the same memory cells. Specifically, the encoding stage $\mathcal{E}_1$ will use only cell levels 0 and 1 to record the indices in its index blocks. Once this stage is over, the index information recorded during $\mathcal{T}_1$ and $\mathcal{E}_1$ is no longer relevant, and the level of *all* the $2(k-1)\mu'$ cells in the $2(k-1)$ index blocks can be raised to 1. Thereafter, provided $q \geqslant 3$, the transition procedure $\mathcal{T}_2$ and the encoding map $\mathcal{E}_2$ can use cell levels 1 and 2 to record the relevant index information in the *same memory cells*. Proceeding in this manner, we can accommodate up to $q-1$ batches of index blocks in $2(k-1)\mu'$ memory cells. We shall refer to this indexing scheme as *stacked binary indexing* and denote the resulting flash code by $\mathbb{C}'$.

**Theorem 3.** *The write deficiency of the flash code $\mathbb{C}'$ defined by the sequence of decoding/encoding maps $\mathcal{D}_0, \mathcal{D}_1, \ldots, \mathcal{D}_{s-1}$ and $\mathcal{E}_0, \mathcal{E}_1, \ldots, \mathcal{E}_{s-1}$ that use stacked binary indexing is at most $O(qk\log k)$ if $q \geqslant \log_2 k$, and at most $O(k\log^2 k)$ otherwise.*

*Proof.* With stacked binary indexing, the number of cell levels wasted in all the $2(k-1)(s-1)$ index blocks is at most

$$2(q-1)(k-1)\left\lceil \frac{s-1}{q-1}\right\rceil \lceil \log_2(k+2)\rceil \quad (7)$$

Although for most values of $k$ and $q$ this is strictly less than (5), all the other terms in (6) are still dominated by (7). ∎

**Remark.** If we need to store $k$ symbols, rather than bits, over an alphabet of size $\ell > 2$, the same flash code can still be used, with an appropriate interface. With the linear womcode of [7], the $\ell$-ary symbols can be represented using $\ell - 1$ bits in such a way that any symbol change corresponds to a single bit transition. The flash code $\mathbb{C}'$ can be now applied as is, and the resulting write deficiency is $O(\max\{q, \log_2 k\ell\}\, k\ell \log k\ell)$.

## V. Flash Codes of Constant Rate

All of our results so far pertain to the case where $n \geqslant k^2$. In this section, we briefly examine the situation where both $k$ and $n$ are large, while $k/n = R$ for some constant $R < 1$. Observe that write deficiency $\delta(\mathbb{C}) = n(q-1) - t$ is *not* an appropriate figure of merit in this situation: a trivial code that guarantees $t = 0$ writes achieves write deficiency $n(q-1) = k(q-1)/R$, which is within a constant factor $2/R$ from the lower bound (1). Thus we will state our results in terms of the guaranteed number of writes $t$ rather than the write deficiency $\delta(\mathbb{C})$.

If $q = 2$, we can easily guarantee $\Omega(n/\log k)$ writes as follows: partition the $n$ cells into blocks of size $\lceil \log_2 k\rceil$ and each time an information bit changes, record its index in the next available block. For $q > 2$, the same method guarantees about $\lfloor n/\log_q k\rfloor = \Omega(n\log q/\log k)$ writes, but we can do better.

Let us partition the $n$ cells into two groups: the *index group* consisting of $n-k$ cells and the *parity group* consisting of $k$ cells. The index group is then subdivided into $m = \lfloor (n-k)/s\rfloor$ blocks, each consisting of $s = \lceil \log_2 k\rceil$ cells. The writing proceeds in $q-1$ phases. During the first phase, every time an information bit changes, its index is recorded in binary (using cell levels 0 and 1) in the next available index block. After $m$ writes, the first phase is over. We then copy the $k$ information bits into the $k$ cells of the parity group, and raise the level of all cells in the index group to 1. The second phase can now proceed using cell levels 1 and 2, and recording changes in information bits relative to the values stored in the parity group. At the end of the second phase, the current values of the $k$ bits are recorded in the parity cells using levels 1 and 2. And so on. This simple coding scheme achieves

$$m(q-1) \;=\; \frac{n(q-1)(1-R)}{\log_2 k} \;=\; \Omega\!\left(\frac{nq}{\log k}\right) \quad (8)$$

writes (where the middle expression ignores ceilings/floors by assuming that $k$ is a power of two and that $n-k$ is divisible by $\log_2 k$). If $q$ is odd and $R \geqslant 0.415$, we can do a little better by using the ternary number system (cell levels $0, 1, 2$) in both the index group and the parity group. In this case, the size of the parity group is $\lceil k/\log_2 3\rceil$ cells and $1 - R$ in (8) can be replaced by $(\log_2 3 - R)/2$. Finally, for all $R \geqslant 0.755$ and $q-1$ divisible by three, the quaternary alphabet is optimal, leading to a factor of $(2-R)/3$ rather than $1 - R$ in (8).